\documentclass[final]{ws-ijmpe}

\newcommand{\tr}{\mbox{Tr}}
\newcommand{\qq}{$\langle q q \rangle$}

\begin{document}

%\preprint{ADP-08-03/T663}

\markboth{P. J.~Moran, D. B.~Leinweber}{Buried treasure in the sand of the QCD vacuum}

%\catchline{}{}{}{}{}
%%%%%%%%%%%%%%%%%%%%%%%%%%%%%%%%%%%%%%%%%%%%%%%%%%%%%%%%%%%%%%%%%%%%

\title{Buried treasure in the sand of the QCD vacuum}

\author{\footnotesize P.J.~Moran, D.B.~Leinweber} \address{Special
  Research Centre for the Subatomic Structure of Matter (CSSM),
  Department of Physics, University of Adelaide, Adelaide SA 5005,
  Australia}

\maketitle

% preprint hack
\vspace{-7cm}
\begin{flushright}
ADP-08-03/T663
\end{flushright}
\vspace{7cm}

%\begin{history}
%  \received{(received date)} \revised{(revised date)}
%  % \accepted{(Day Month Year)} \comby{(xxxxxxxxxx)}
%\end{history}

\begin{abstract}
  The short-range structure of the $2+1$ flavour QCD vacuum is studied
  through visualisations of the topological charge density.  Of
  particular interest is a new Gaussian weighted smearing algorithm
  which is applied to the rough topological charge density to disclose
  underlying long range structure.  The results provide support for
  the view of the QCD vacuum as a sandwich of sign-alternating sheets
  of charge, with a long-range structure hidden beneath.
\end{abstract}

\section{Introduction}

The topological charge density of the QCD vacuum is most often
obtained
through either of two possible approaches. The first is through the
gluonic definition,
\begin{equation}
  \label{eq:gtopqx}
  q(x) = \frac{g^2}{32 \pi^2} \epsilon_{\mu\nu\rho\sigma} \tr [
  F_{\mu\nu}(x) F_{\rho\sigma}(x) ] \,,
\end{equation}
in combination with some cooling\cite{cool1,cool2,cool3} or
smearing\cite{smear1,smear2,oimpstout,fivesweeps}
algorithm. Significant
work has been invested into refining the relevant cooling/smearing
operators in order to reduce the effects of discretisation errors
during the cooling/smearing process.\cite{oimpstout,imp1,imp2}

Another approach is through the fermionic
definition,\cite{fermqx1,fermqx2}
\begin{equation}
  \label{eq:ftopqx}
  q(x) = - \mathrm{tr} \left[ \gamma_5 \left( 1 - \frac{a}{2} D(0;x,x)
    \right) \right] \,,
\end{equation}
defined with the overlap Dirac operator,\cite{overlap1,overlap2}
\begin{equation}
  \label{eq:diracop}
  D(0) = \frac{\rho}{a}\left( 1 + D_W/\sqrt{D_W^{\dagger} D_W} \right)
  = \frac{\rho}{a}\left( 1 + \gamma_5 \mathrm{sgn} (H_W) \right) \,,
\end{equation}
where $H_W = \gamma_5 D_W$ and $D_W$ is a Wilson-type Dirac
operator. This definition satisfies the
Atiyah-Singer index theorem
\begin{equation}
  Q = n_{-} - n_{+} \,,
  \label{eq:Atiyah-Singer}
\end{equation}
and hence, will always return an integer total topological charge $Q$. The
overlap Dirac operator is far more computationally expensive to
compute than using either cooling or smearing, however advances in
techniques and computing power have enabled some interesting studies.

\begin{figure}
  \centering
  \begin{tabular}{cc}
    \includegraphics[width=0.48\textwidth,angle=0]{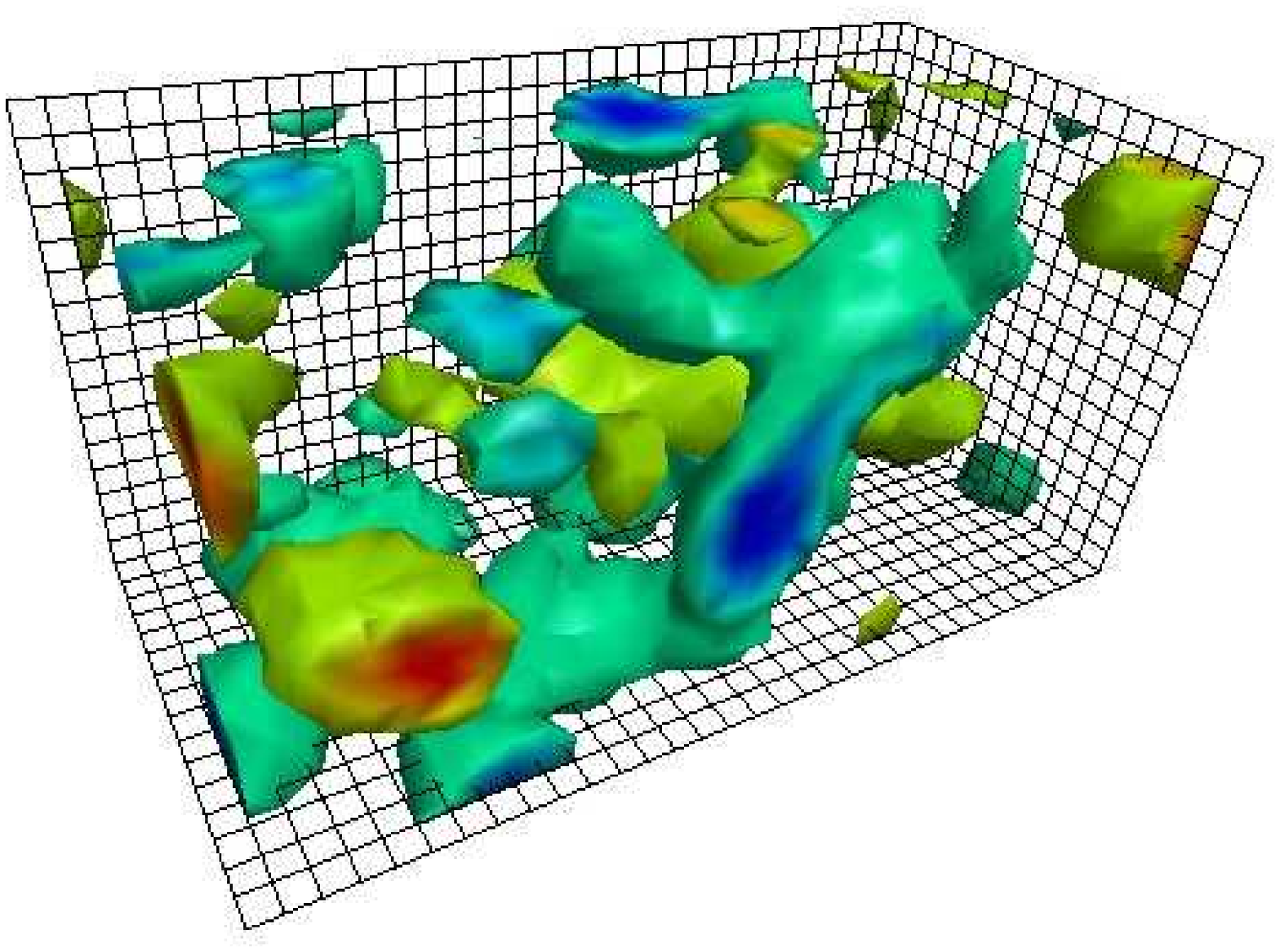}&
    \includegraphics[width=0.48\textwidth,angle=0]{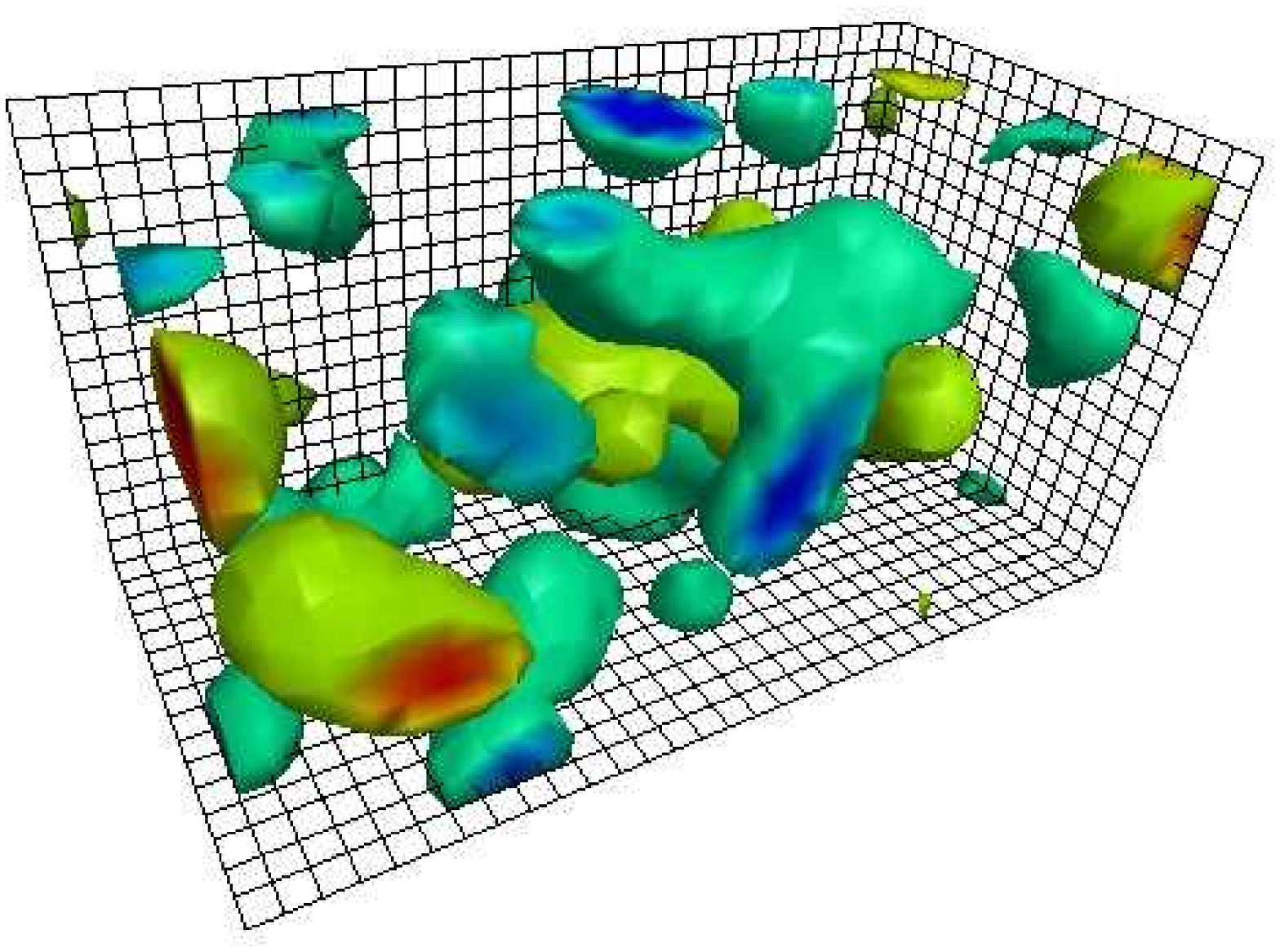}
  \end{tabular}
  \caption{The best match between the overlap topological charge
    density $q(x)$ [left] and 
    the over-improved $q(x)$ [right] for a $\lambda_{\rm cut} =
    634$~MeV, which gives $n_{sw} = 48$ as reported by Ilgenfritz {\it
    et al.}\protect\cite{fermvsglue}}
  \label{fig:qxmatch634}
\end{figure}

Recently, Ilgenfritz {\it et al.}\cite{fermvsglue} have demonstrated how
the use of the
spectral representation for the Dirac
operator,\cite{spectral1,spectral2} 
\begin{equation}
  \label{eq:truncateddirac}
  q_{\lambda_{cut}}(x) = - \sum_{|\lambda|<\lambda_{cut}} 
  \left( 1 - \frac{\lambda}{2} \right) \psi_{\lambda}^{\dagger}(x)
  \gamma_5 \psi_{\lambda}(x) \,,
\end{equation}
enables one to calibrate the number of over-improved stout-link
smearing\cite{oimpstout} sweeps to a given spectral cut-off,
$\lambda_{\rm cut}$. In
particular, five 
sweeps of smearing was matched with the full overlap density, and it was
found that a $\lambda_{\rm cut}$ of $634$~MeV corresponds to
approximately 45 sweeps of over-improved stout-link smearing, as
seen in Fig.~\ref{fig:qxmatch634}.

The strong correlation between the different approaches justifies the
use of over-improved smearing for efficient, large scale studies
of QCD vacuum structure. For example, using five sweeps of smearing
to generate a negative $\langle q q \rangle \equiv \langle q(x) q(0)
\rangle$ correlator,\cite{fivesweeps} the effects of dynamical
fermions on the 
short-range structure of the vacuum have been studied.\cite{dynvac} In
the
following, we use over-improved stout-link smearing to further probe
the UV structure of the QCD vacuum. 

\section{Short distance structure}

The negative behaviour of the \qq~correlator\cite{negqq} suggests a
sign-alternating sheet-like topological structure exists in the QCD
vacuum.\cite{sheet1,sheet2}

Using a $28\times96$, dynamical lattice with $a m_{u,d} = 0.0062$, $a
m_{s} = 0.031$, generously
provided by the MILC collaboration,\cite{milc1,milc2} we use five
sweeps of
over-improved smearing in combination with a 3-loop improved
topological charge operator\cite{imp2} to study the short-range vacuum
structure.

\begin{figure}[t]
  \centering
  \begin{tabular}{cc}
    \includegraphics[width=0.48\textwidth,angle=0]{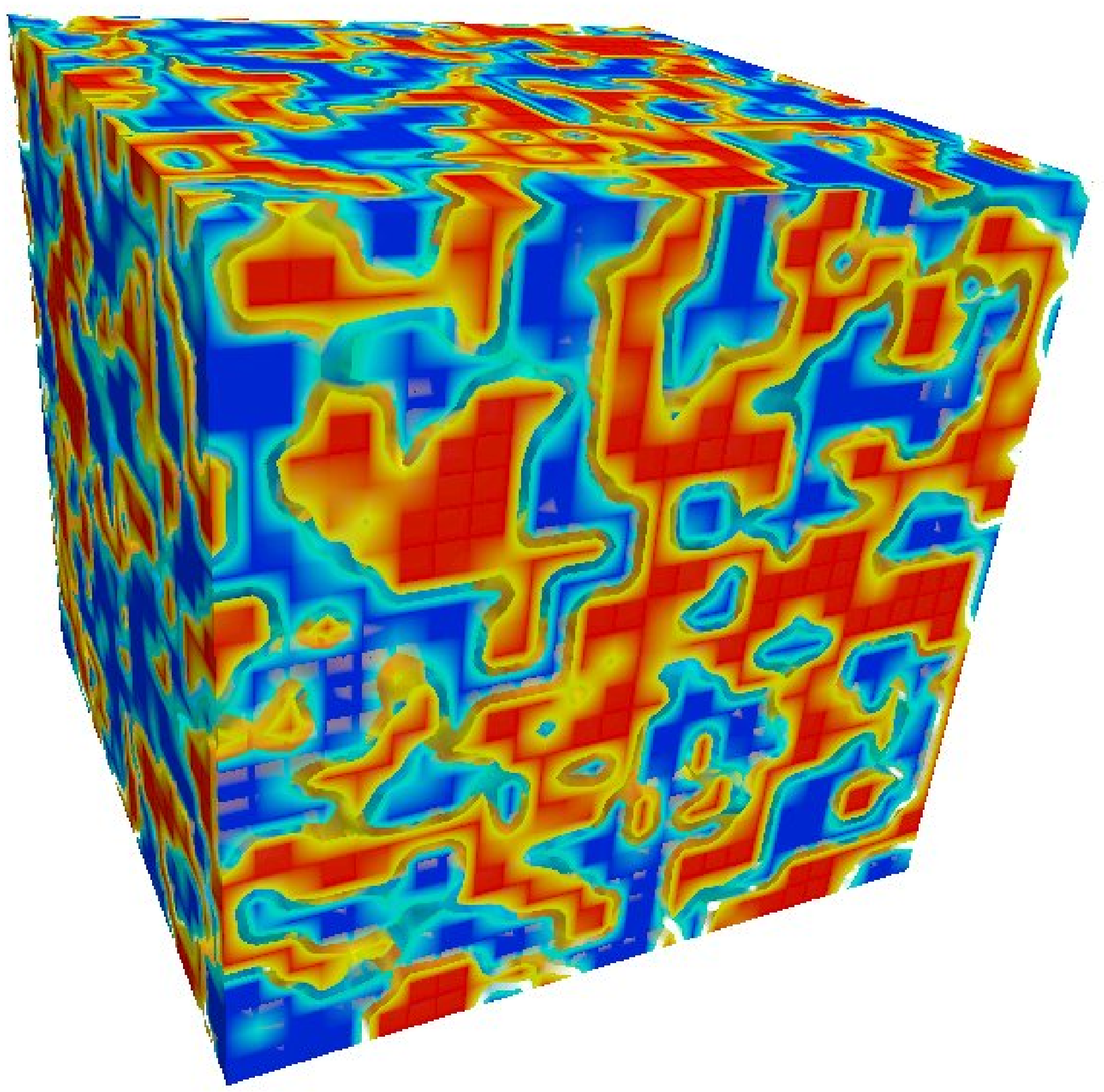}&
    \includegraphics[width=0.48\textwidth,angle=0]{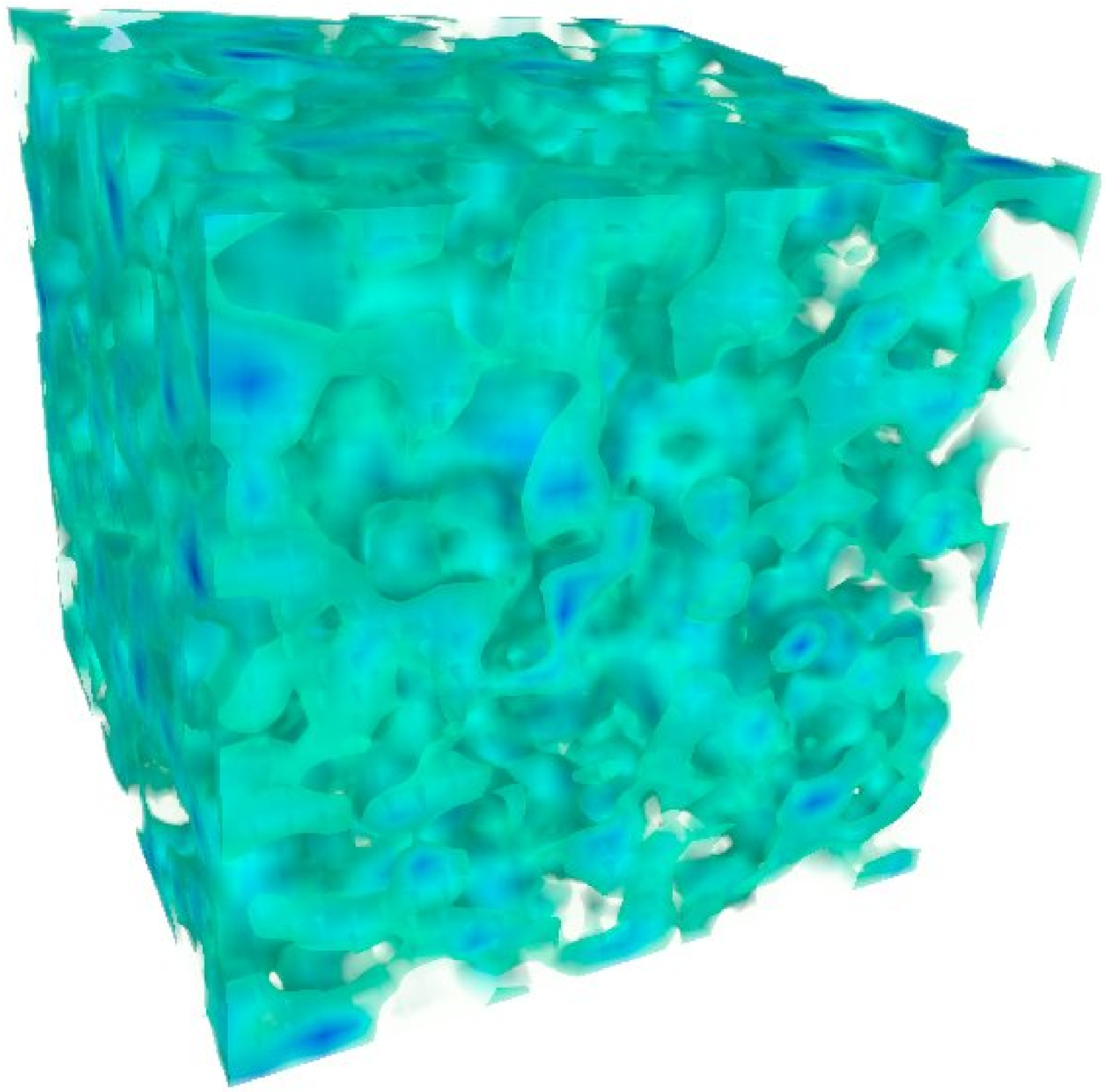}
  \end{tabular}
  \caption{[left] The short-distance sheet structure of the vacuum is
    clearly apparent after five sweeps of over-improved stout-link
    smearing. Negative charge density is green to blue, and positive charge
    density is yellow to red. [right] The same data, this time with the positive
    charge removed and the magnitude of the negative charge
    shown through the strength of the blue colouring.}
  \label{fig:sheetandpartial}
\end{figure}

In Fig.~\ref{fig:sheetandpartial} we present two visualisations of
the topological charge density. In the left graphic we have plotted the
full topological charge density range including values approaching
zero, colouring negative charge green to blue
and positive charge yellow to red. This approach reveals the proposed sheet-like
structure of the vacuum.\cite{sheet1} In the right graphic we render only the
negative charge, colouring from light to dark the regions of
weakest to strongest charge, illustrating the structure that exists
within the sheets.

\begin{figure}[t]
  \centering
  \begin{tabular}{cc}
    \includegraphics[width=0.48\textwidth,angle=0]{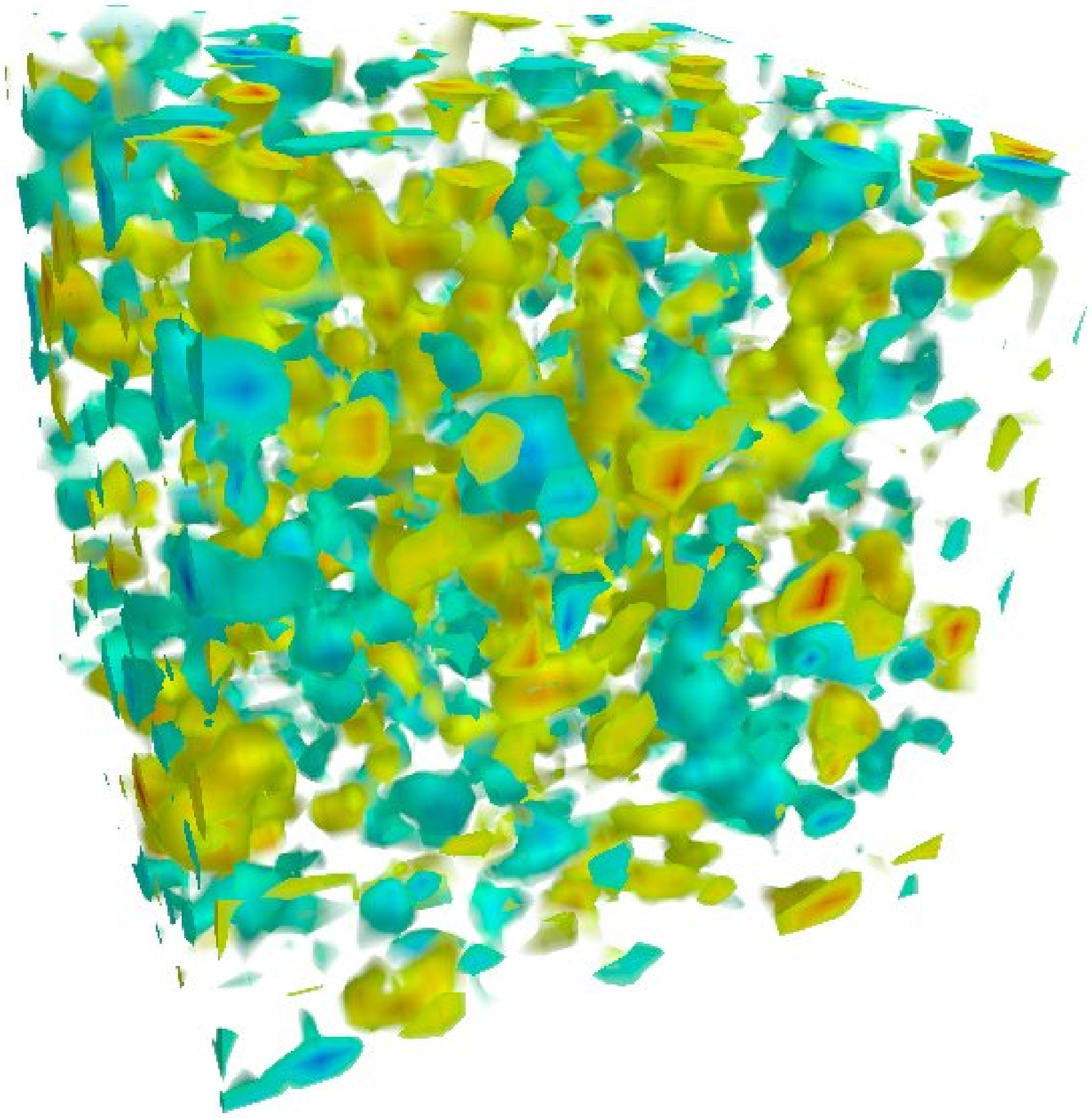}&
    \includegraphics[width=0.48\textwidth,angle=0]{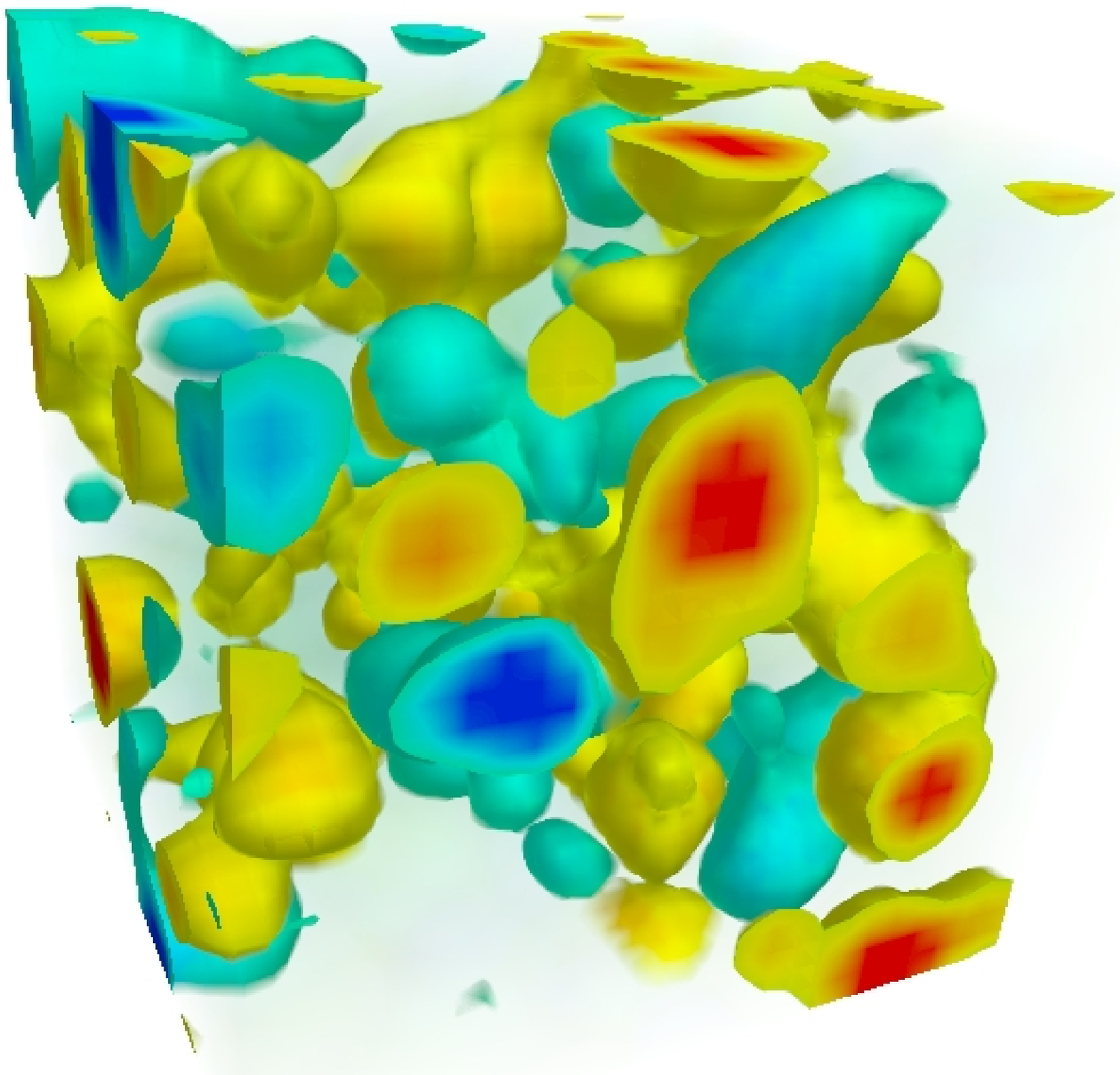}
  \end{tabular}
  \caption{[left] Placing a high cutoff on $q(x)$ such that only the
    most intense regions of charge are seen presents a picture of the
    QCD vacuum which resembles the sand of one of Australia's many
    fine beaches.  The grains of sand will diminish in size as the
    continuum limit is approached.  [right] The topological charge
    after 45 sweeps of over-improved stout-link smearing.}
  \label{fig:sandandsw45}
\end{figure}
Raising the cutoff threshold for $q(x)$ so that only the most intense
regions of charge are shown presents a different
view of the vacuum, as seen in the left graphic of
Fig.~\ref{fig:sandandsw45}. Here the
vacuum appears to have a granular, sand-like structure. Ilgenfritz
has investigated this idea of clustering deeply.\cite{clustering} The
right plot
of Fig.~\ref{fig:sandandsw45} shows the topological charge density
after 45 sweeps of over-improved stout-link smearing, where we see the
familiar instanton-like 
lava lamp structure. It is tempting to try and find
similarities between the two plots, however the vast number of sandy
objects prohibits us from doing so.

\begin{figure}[t]
  \centering
  \begin{tabular}{cc}
    \includegraphics[width=0.48\textwidth,angle=0]{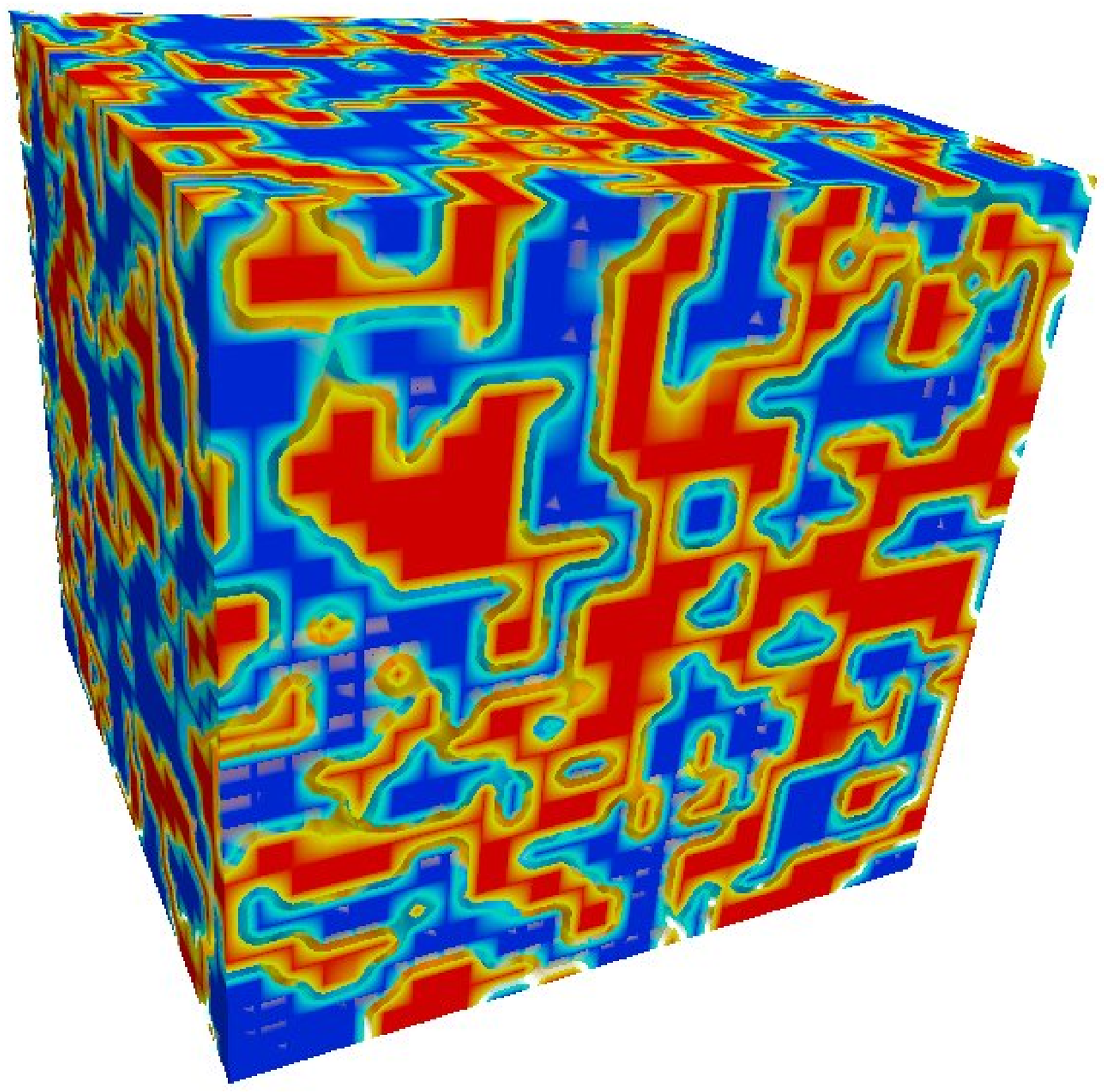}&
    \includegraphics[width=0.48\textwidth,angle=0]{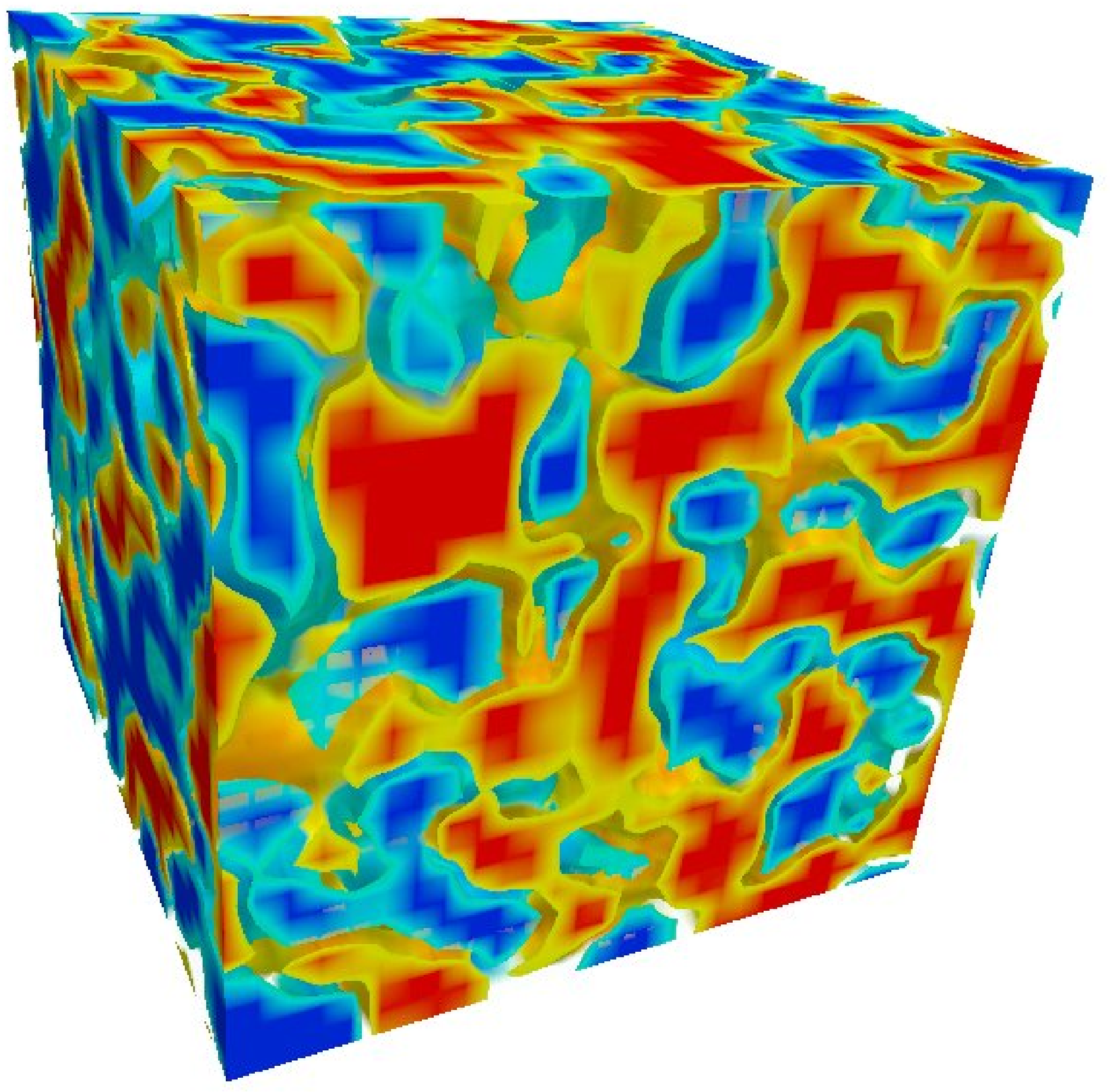}\\
    \includegraphics[width=0.48\textwidth,angle=0]{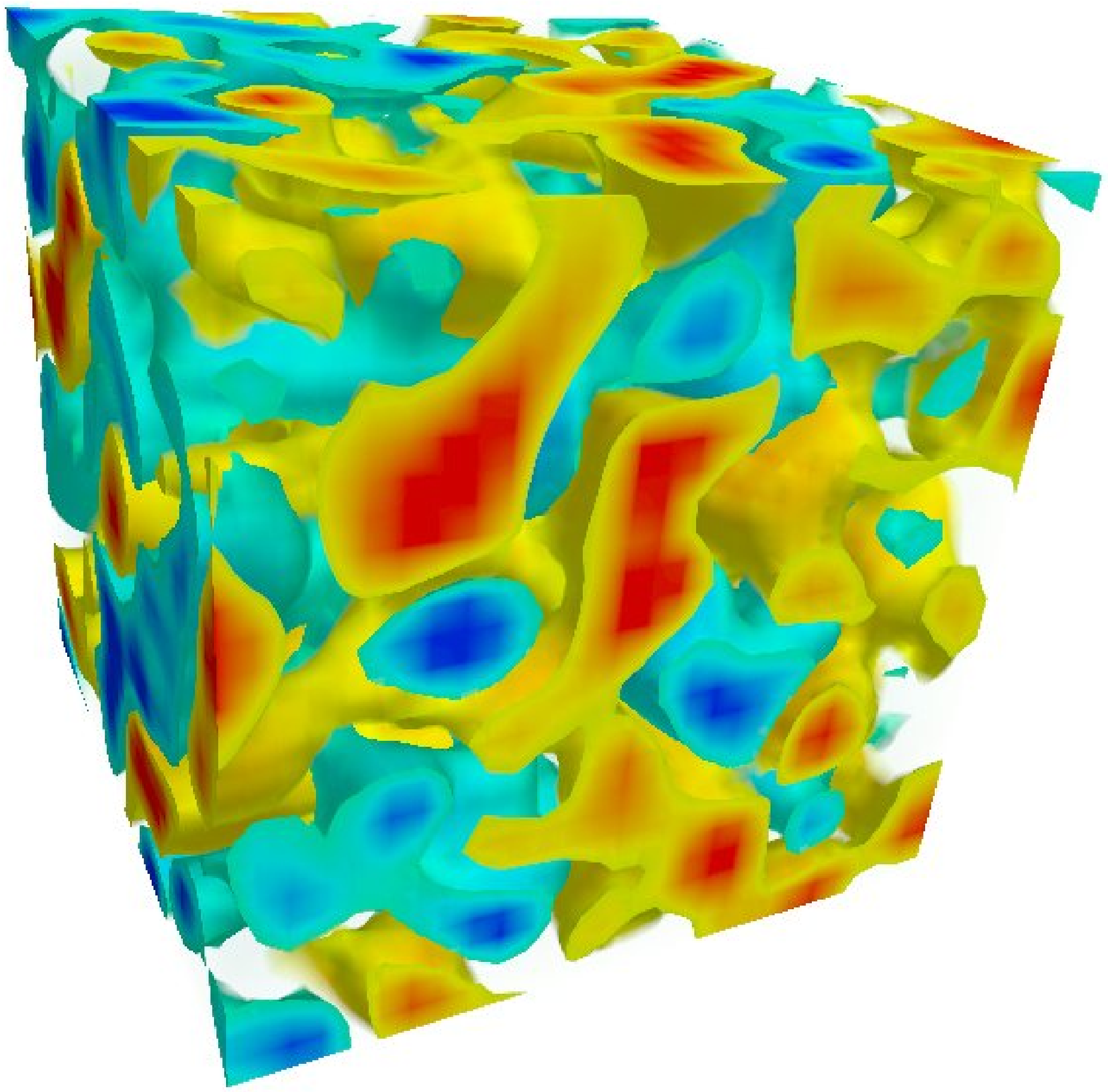}&
    \includegraphics[width=0.48\textwidth,angle=0]{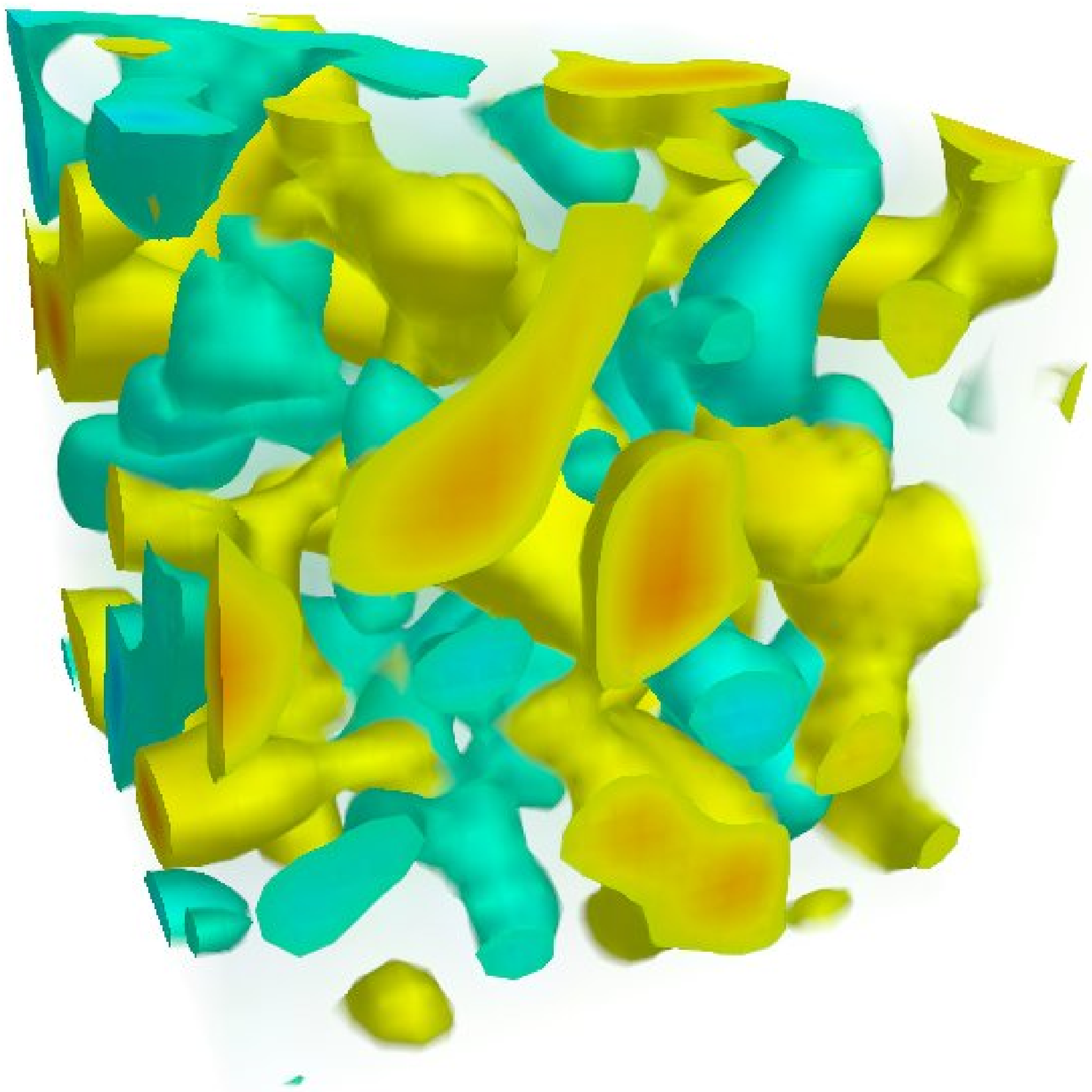}
  \end{tabular}
  \caption{The evolution of $q(x)$ obtained from five sweeps of
    over-improved stout-link smearing under Gaussian smoothing. The
    values of $\sigma$ used were 0.25 [top-left], 0.75 [top-right],
    1.25 [bottom-left], and 1.75 [bottom-right].}
  \label{fig:bigsigma}
\end{figure}

In order to draw some comparisons between the two different sheet
and lava pictures of
the QCD vacuum we need some
method of averaging the sheet structure of
Fig.~\ref{fig:sheetandpartial}. For this, we define a Gaussian
smoothing operation to act on the topological charge density
itself. Given $q(x)$ for some gauge field, simultaneously update each
point $x_0$ on the lattice according to,
\begin{equation}
  q(x_0) = \frac{1}{2\pi\sigma^2} \sum_{x}
  e^{-r^2/2\sigma^2} q(x) \,,
\end{equation}
where $r$ is the Euclidean distance between $x$ and $x_0$, and
$\sigma$ is the
standard deviation of the Gaussian distribution in lattice units.

The evolution of the topological charge density of
Fig.~\ref{fig:sheetandpartial}
under Gaussian smoothing with increasing $\sigma$ is shown in
Fig.~\ref{fig:bigsigma}. The resulting effect appears to be
quite similar to that seen in previous cooling/smearing animations.

\begin{figure}
  \centering
  \begin{tabular}{cc}
    \includegraphics[width=0.48\textwidth,angle=0]{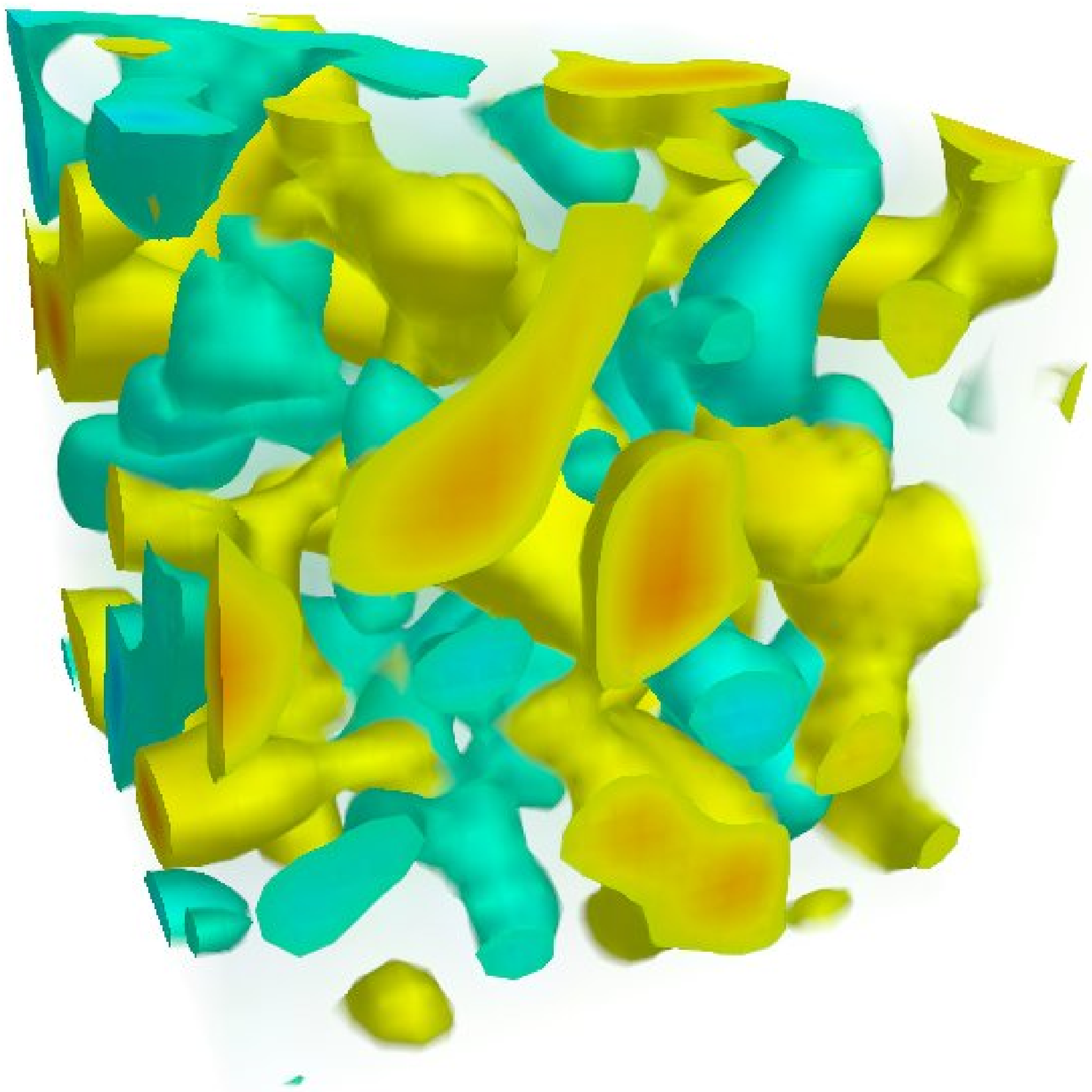}&
    \includegraphics[width=0.48\textwidth,angle=0]{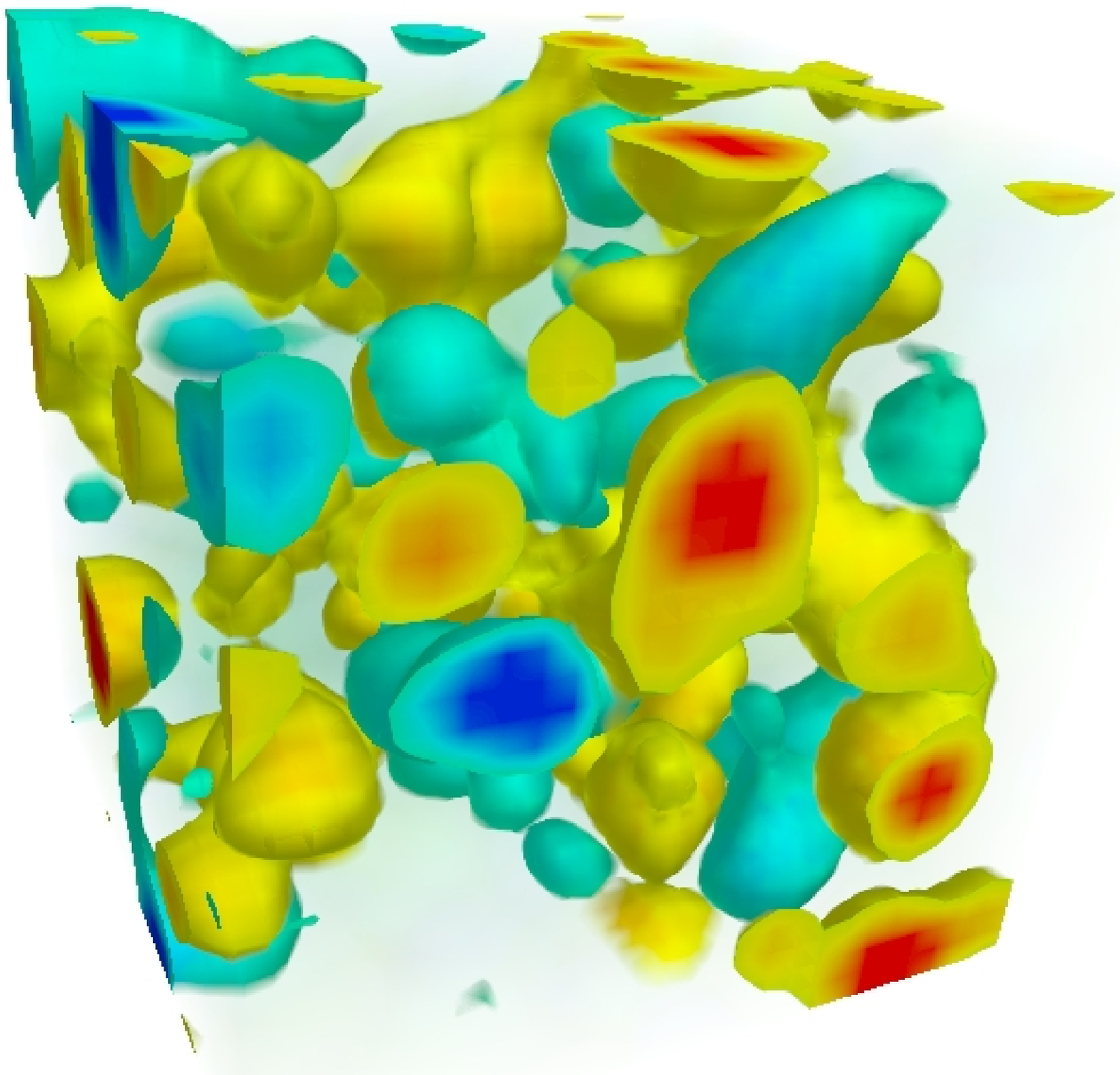}
  \end{tabular}
  \caption{[left] The Gaussian smoothed topological charge density
    after five smearing sweeps, using $\sigma = 1.75$. [right] The
    topological charge density after 45 smearing sweeps. The two
    pictures are far from identical, however it is remarkable that
    a few common features are present.}
  \label{fig:sigvs45}
\end{figure}

In Fig.~\ref{fig:sigvs45} there is a side by side comparison of the
Gaussian smoothed charge density with $\sigma=1.75$ and the
topological charge density
obtained after 45 sweeps of smearing. Recall that the Gaussian
smoothed charge density was generated from the topological charge
density after only
five smearing sweeps. Although the resulting densities are certainly
not identical, they still share many common
features.

This kind of calculation suggests that the UV sheet structure of the
vacuum is embedded on top of an underlying long-range structure. With
the long-range structure revealed through the application of a
cooling/smearing operator or through a truncated overlap Dirac
operator. This kind of idea has been discussed previously by P. de
Forcrand.\cite{forcrand}

\section{Conclusion}

\begin{figure}
  \centering
  \includegraphics[width=0.5\textwidth,height=2.1cm,angle=0]{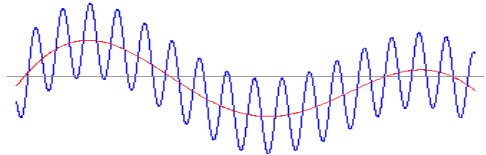}
  \caption{An example of how UV fluctuations could be superimposed on a
  deeper long-range structure, as suggested by P. de
  Forcrand.\protect\cite{forcrand}}
  \label{fig:shortlongrange}
\end{figure}

Using a Gaussian smoothing procedure the rough topological charge
density
obtained from five sweeps of over-improved stout-link smearing was
compared with that obtained from 45 sweeps. Modest agreement was found
using a Gaussian standard deviation of $\sigma = 1.75$. These results
suggest that the QCD vacuum consists of a sandwich of high-energy
fluctuations, with a long-distance structure hidden beneath. This is
represented graphically in Fig.~\ref{fig:shortlongrange}.

\section*{Acknowledgements}

We thank the Australian Partnership for Advanced Computing (APAC) and
the South Australian Partnership for Advanced Computing (SAPAC) for
generous grants of supercomputer time which have enabled this
project. This work is supported by the Australian Research Council.


\begin{thebibliography}{0}

\bibitem{cool1} B. Berg, {\it Phys. Lett.} {\bf B104}, (1981) 475.
\bibitem{cool2} M. Teper, {\it Phys. Lett.} {\bf B162}, (1985) 357.
\bibitem{cool3} E.-M. Ilgenfritz et al., {\it Nucl. Phys.} {\bf B268},
  (1986) 693.
\bibitem{smear1} M. Albanese et al., {\it Phys. Lett.} {\bf B192},
  (1987) 163.
\bibitem{smear2} F. D. R. Bonnet et al., {\it Phys. Rev.} {\bf D65}, (2002)
  114510.
\bibitem{oimpstout} P. J. Moran and D. B. Leinweber, {\it Phys. Rev.}
  {\bf D77}, (2008), 094501.
\bibitem{fivesweeps} P. J. Moran and D. B. Leinweber, {\it PoS} {\bf LATTICE
    2007}, (2007) 383.
\bibitem{imp1} P. de Forcrand et al., {\it
  Nucl. Phys. Proc. Suppl.} {\bf B47}, (1996), 777.
\bibitem{imp2} S.O. Bilson-Thompson et al., {\it Ann. Phys. (N.Y.)}
  {\bf 304}, (2003), 1.
\bibitem{fermqx1} P. Hasenfratz et al., {\it Phys. Lett.} {\bf B427},
  (1998) 125.
\bibitem{fermqx2} F. Niedermayer, {\it Nucl. Phys. Proc. Suppl.} {\bf
  73}, (1999) 105.
\bibitem{overlap1} H. Neuberger, {\it Phys. Lett.} {\bf B417}, (1998)
  141.
\bibitem{overlap2} H. Neuberger, {\it Phys. Lett.} {\bf B427}, (1998)
  353.
\bibitem{fermvsglue} E.-M. Ilgenfritz et al., {\it Phys. Rev.} {\bf D77},
  (2008) 074502.
\bibitem{dynvac} P. J. Moran and D. B. Leinweber, arXiv:0801.2016
  [hep-lat].
\bibitem{spectral1} I. Horvath et al., {\it Phys. Rev.} {\bf D67}, (2003)
  011501.
\bibitem{spectral2} Y. Koma et al., {\it PoS} {\bf LAT2005}, (2006) 300.

\bibitem{clustering} E.-M. Ilgenfritz et al., {\it Phys. Rev.} {\bf D76},
  (2007) 034506.

\bibitem{negqq} I. Horvath et al., {\it Phys. Lett.} {\bf B617}, (2005)
  49.
\bibitem{sheet1} I. Horvath et al., {\it Phys Rev.} {\bf D68}, (2003)
  114505.
\bibitem{sheet2} I. Horvath et al., {\it Phys. Lett.} {\bf B612},
  (2005) 21.
\bibitem{milc1} Aubin et al., {\it Phys. Rev.} {\bf D70}, (2004) 094505.
\bibitem{milc2} Bernard et al., {\it Phys. Rev.} {\bf D64}, (2001)
  054506.
\bibitem{forcrand} P. de Forcrand, {\it AIP Conf. Proc.} {\bf 892},
  (2007) 29.

\end{thebibliography}
\end{document}